\documentclass[
    ,final            % use final for the camera ready runs
  ]
  {aipproc}

\newcommand{\grado}{^{\circ}}

\layoutstyle{6x9}

\begin{document}

\title{Dark Matter in Draco: new considerations of the expected gamma flux in IACTs\footnote{Talk presented by M.A.S.C. at The Dark Side of the Universe International Workshop, Madrid, Spain, 20-24 June 2006}}


\author{Miguel A. S\'anchez-Conde}{
  address={Instituto de Astrofisica de Andalucia (CSIC), E-18008, Granada, Spain}
}

\author{Francisco Prada}{
  address={Instituto de Astrofisica de Andalucia (CSIC), E-18008, Granada, Spain}
}

\author{Ewa L. {\L}okas}{
  address={Nicolaus Copernicus Astronomical Centre, Bartycka 18, 00-716 Warsaw, Poland\\ e-mail: masc@iaa.es; fprada@iaa.es; lokas@camk.edu.pl}
}

\begin{abstract}
  A new revision of the gamma flux that we expect to detect in Imaging Atmospheric Cherenkov Telescopes (IACTs) from SUSY dark matter annihilation in the Draco dSph is presented using the dark matter density profiles compatible with the latest observations. This revision takes also into account the important effect of the Point Spread Function (PSF) of the Cherenkov telescope. We show that this effect is crucial in the way we will observe and interpret a possible signal profile in the telescope. Given these new considerations, some light can be shed on the recent signal excess reported by the CACTUS experiment.
\end{abstract}

\maketitle

%%%%%%%%%%%%%%%%%%%%%%%%%%%%%%%%%%%%%%%%%%%%
%% MAINMATTER
%%%%%%%%%%%%%%%%%%%%%%%%%%%%%%%%%%%%%%%%%%%%

\section{Introduction}

Nowadays, the indirect search for supersymmetric (SUSY) dark matter is possible by means of the new Imaging Atmospheric Cherenkov Telescopes (IACTs). This search is based on the detectability of gamma rays coming from the annihilation of SUSY dark matter particles that takes place in those places in the Universe where the dark matter density is high enough. IACTs in operation like MAGIC or HESS, or in the near future the GLAST satellite or GAW, will play a very important role in this kind of SUSY dark matter searches.

The expected total number of continuum $\gamma$-ray photons received per unit time and per unit area, from a circular aperture on the sky of width $\sigma_{\rm t}$ (which represents the resolution of the telescope) observing at a given direction $\Psi_0$ relative to the centre of the dark matter halo is given by:

\begin{equation}
F(E>E_{\rm th})=\frac{1}{4\pi} {f_{SUSY}} \cdot U(\Psi_0), 
\label{eq1}
\end{equation}
$${f_{SUSY}}= \frac{N_{\gamma} \left<\sigma v\right>}{2 m_\chi^2}, \qquad U(\Psi_0)=\int J(\Psi)B(\Omega)d\Omega,$$

\noindent where the factor $f_{SUSY}$ encloses all the particle physics, and the factor $U(\Psi_0)$ involves all the astrophysical properties (such as the dark matter distribution and geometry considerations). This astrophysical factor also accounts for the beam smearing, where:

\begin{equation}
J(\Psi) =  \int_{l.o.s} \rho_{dm}^2(r) \, dl , \qquad dl = \pm r dr/{\sqrt{r^2-d_{\odot}^2 \sin^2\Psi}}
\label{eq2}
\end{equation}
is the integral of the line-of-sight of the square of the dark matter density along the direction $\Psi$, and $B(\Omega)d\Omega$ is the Gaussian beam of the telescope:

\begin{equation}
B(\Omega) d\Omega  =  \exp\left[ -\frac{\theta^2}{2\sigma_t^2}\right] 
sin\theta \, d\theta d\varphi .
\label{eq3}
\end{equation}

\noindent This last factor, commonly known in the astrophysical community as the Point Spread Function (PSF) of the instrument, plays a very important role in the way we will "see" a possible signal in the telescope. However, most of previous work in the literature did not take into account its effect (except \citep{Prada04}).

A very important question concerning the indirect search of SUSY dark matter is where to search. Because of the fact that the factor $U(\Psi_0)$ gamma flux is proportional to the square of the dark matter density, we will need to point our IACTs telescopes to places with a high concentration of dark matter. In principle, the best option seems to be the Galactic Center (GC), since it satisfies this condition and also it is very near compared to other potential targets. However, the GC is a very crowded region, which makes it difficult to discriminate between a possible $\gamma$-ray signal due to dark matter annihilation and other astrophysical sources.

There are also other possible targets with high dark matter density in relative proximity from us, which are not plagued by the problem of the GC, e.g. the Andromeda galaxy, the dwarf spheroidal (dSph) galaxies - most of them satellites of the Milky Way- or even huge cluster of galaxies (e.g. Virgo). dSph galaxies represent a good option, since they show very high mass to light ratios, and at least six of them are nearer than 100 kpc from the GC (Draco, LMC, SMC, CMa, UMi and Sagittarius). Draco is the dSph with strongest observational constraints, located at 80 kpc. This fact is very important if we really want to make a realistic prediction of the expected $\gamma$-ray flux. Moreover, recently the CACTUS collaboration reported a possible gamma-ray excess from Draco, which makes a detailed study of this dSph even more attractive and necessary.

\section{Dark matter distribution in Draco}

In our modelling of Draco we used the sample of 207 Draco stars with measured line-of-sight velocities originally considered as members by \citep{wkeg}. We used a rigorous method of removal of possible interlopers originally proposed by \citep{hartog} and applied to galaxy clusters. The method relies on calculating the maximum velocity available to the members of the galaxy assuming that they are on circular orbits or infalling into the structure. The method was shown to be the most efficient among many others for interloper removal recently tested on cluster-size simulated dark matter haloes by \citep{wlgm}. Its applicability and efficiency in the case of dSph galaxies was demonstrated by \citep{klim}. 

 We assumed that the dark matter distribution in Draco can be approximated by the formula:
\begin{equation}    \label{kazantzidis}
    \rho_{\rm d}(r) = C r^{-\alpha} {\rm exp} \left( -\frac{r}{r_{\rm b}} \right)
\end{equation}
proposed by \citep{kmmds}, which was found to fit the density distribution of a simulated
dwarf dark matter halo stripped during its evolution in the potential of a giant galaxy.
\citep{kmmds} found that the halo, which initially had a NFW distribution, preserves the cusp
in the inner part (so that $\alpha=1$ fits the final remnant very well) but develops an
exponential cut-off in the outer parts. Here we will consider two cases, the profile with a
cusp $\alpha=1$ and a core $\alpha=0$. It remains to be investigated what scenarios could lead to
such core profiles.

\begin{figure}
  \includegraphics[height=.3\textheight]{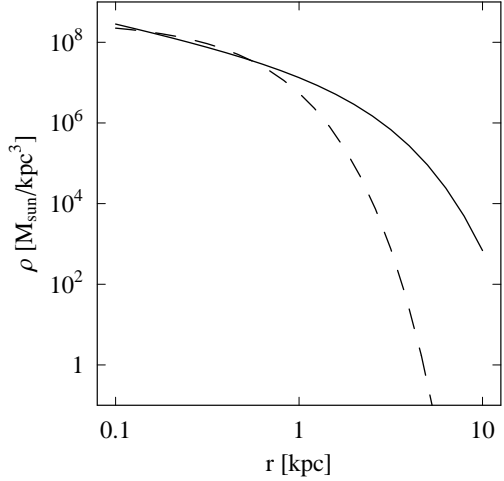}
  \caption{The best-fitting dark matter density profiles for Draco with a cusp (solid line) and
a core (dashed line).}  \label{density194}
\end{figure} 

We modelled the velocity dispersion and kurtosis profiles calculated from the data solving the Jeans equations with dark matter profiles given by Eq.(4). The best-fitting solutions in the case of a cusp and a core profile are given in Table 1. $M_{\rm D}/M_{\rm S}$ is the ratio of the total dark matter mass to total stellar mass, $r_{\rm b}/R_{\rm S}$ the break radius of Eq.(\ref{kazantzidis}) in units of the S\'ersic radius of the stars and $\beta$ the anisotropy parameter of the stellar orbits.

\begin{table}
\begin{tabular}{ccccccccc}
profile &  & $M_{\rm D}/M_{\rm S}$ & & $r_{\rm b}/R_{\rm S}$ & & $\beta$  & & $\chi^2/N$  \\
\hline
cusp    &  & 830  & &  7.0 & & $-0.1$ & &  8.8/9   \\
core    &  & 185  & &  1.4 & & 0.06  &  &  9.5/9   \\
\hline
\end{tabular}
\caption{Best-fitting parameters for the dark matter profiles with a cusp ($\alpha=1$) and a core ($\alpha=0$). The last column gives the goodness of fit measure $\chi^2/N$.}
\label{parameters}
\end{table}

Fig.~\ref{density194} shows the best-fitting dark matter density profiles in the case of the cusp
(solid line) and the core (dashed line). As we can see, both density profiles are similar
up to about 1 kpc, where they are constrained by the data. The reason
for very different values of the break radius $r_{\rm b}$ in both cases is the
following. The kurtosis is sensitive mainly to anisotropy and it forces $\beta$
to be close to zero in both cases. However, to reproduce the velocity
dispersion profile with $\beta \approx 0$ the density profile has to be steep
enough. In the case of the core it means that the exponential cut-off has to
occur for rather low radii, which is what we see in the fit. The cusp
profile does not need to steepen the profile so much so it is much more
extended and its total mass is much larger.

\section{The role of the PSF}

\begin{figure}[!hb]
  \includegraphics[height=.3\textheight]{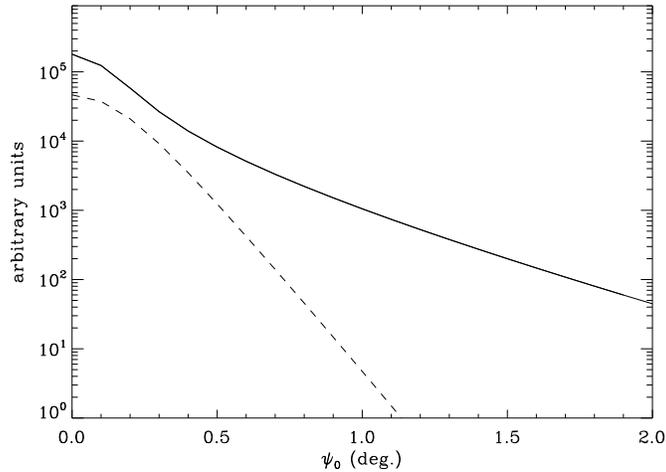}
  \caption{Draco flux predictions for the core (dashed line) and cusp (solid line) density profiles. A PSF of $0.1\grado$ was used}  \label{cuspcorepsf01}
\end{figure}

In order to compute the expected gamma flux, we need to calculate the value of the astrophysical factor $U(\psi_0)$, given in Eq.(\ref{eq1}), for the core ($\alpha=0$) and cuspy ($\alpha=1$) density profiles as given by Eq.(\ref{kazantzidis}) with the parameters listed in Table \ref{parameters}. The results are plotted in Figure \ref{cuspcorepsf01}, where we used a PSF of $0.1\grado$ (which is the typical PSF for an IACT like MAGIC or HESS). As we can see, it is possible to distinguish between both density profiles thanks to a different and characteristic shape in each case.

\begin{figure}
  \includegraphics[height=.3\textheight]{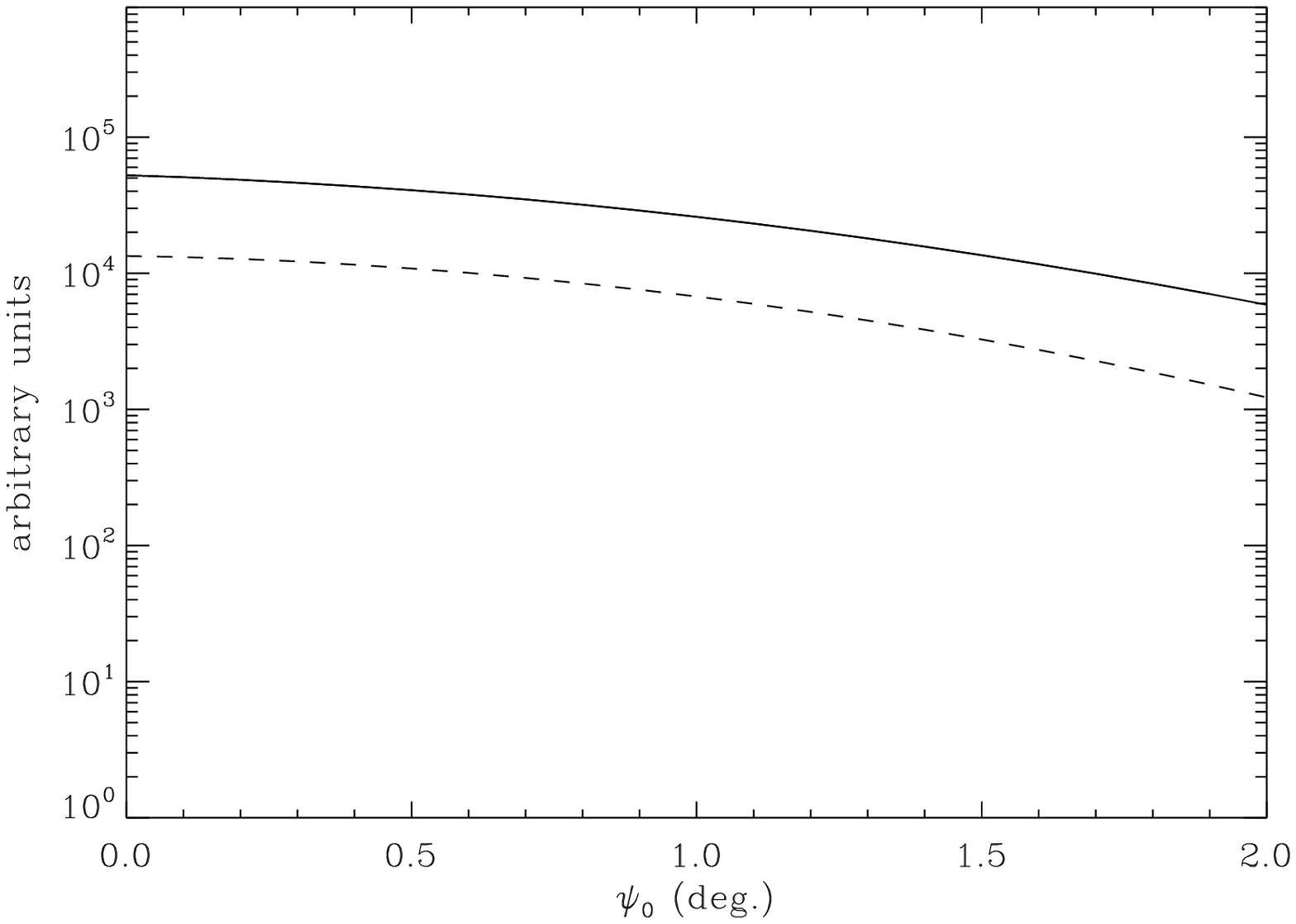}
  \caption{Same as Figure \ref{cuspcorepsf01} but for a PSF$=1\grado$.}
  \label{cuspcorepsf1}
\end{figure}

To illustrate the PSF effect on the shape of the flux profile measured with our IACT, in Figure \ref{cuspcorepsf1} we show the same as in Figure \ref{cuspcorepsf01}, but here for a PSF$=1\grado$ (the PSF of the CACTUS experiment). It is clear that, although we use different dark matter density profiles, a worse PSF makes both resultant flux profiles indistinguishable. According to these results, however, one may think that we could distinguish them by means of the value of the flux near the centre, i.e. for small values of $\psi_0$. However, this is not possible in practice due to the large uncertainties, which come mainly from the particle physics (uncertainties around three or four orders of magnitude in most cases, see \citep{Prada04}).

\begin{figure}
  \includegraphics[height=.3\textheight]{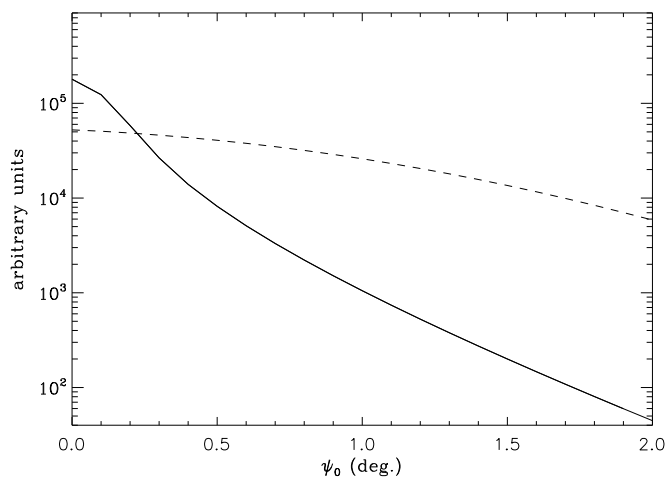}
  \caption{Draco flux predictions for the cusp density profile, and for two different PSFs. Solid line corresponds to PSF=$0.1\grado$ and  dashed line to PSF=$1\grado$.}  
  \label{cusppsfs}
\end{figure}

For the same dark matter density profile, a worse PSF flattens the flux profile. It can be clearly seen in Figure \ref{cusppsfs}, where we plot the flux predictions for the cuspy density profile, and for two different values of the PSF ($0.1\grado$ and $1\grado$). It is also possible to see the same effect if we make a comparison between Figures \ref{cuspcorepsf01} and \ref{cuspcorepsf1}.

\section{A discussion of CACTUS results}

CACTUS is a ground based $\gamma$-ray telescope located in California. The experiment is sensitive to $\gamma$-rays above 50 GeV and it has an effective area of about $50000~m^2$. It was first designed for solar observations and not for $\gamma$-ray astronomy. Because of this fact, it has a poor PSF of around $1\grado$. Recently, the CACTUS collaboration reported a $\gamma$-ray excess from Draco (\citep{marleau}). In \citep{profumo&kamion}, the CACTUS data were superimposed on different flux profiles, each of them related to different models of the dark matter density profile. However, we must note here that their flux estimations were computed without taking into account the important role of the PSF.

\begin{figure}
  \begin{minipage}[t]{0.5\linewidth}
    \centering
    \includegraphics[height=.25\textheight]{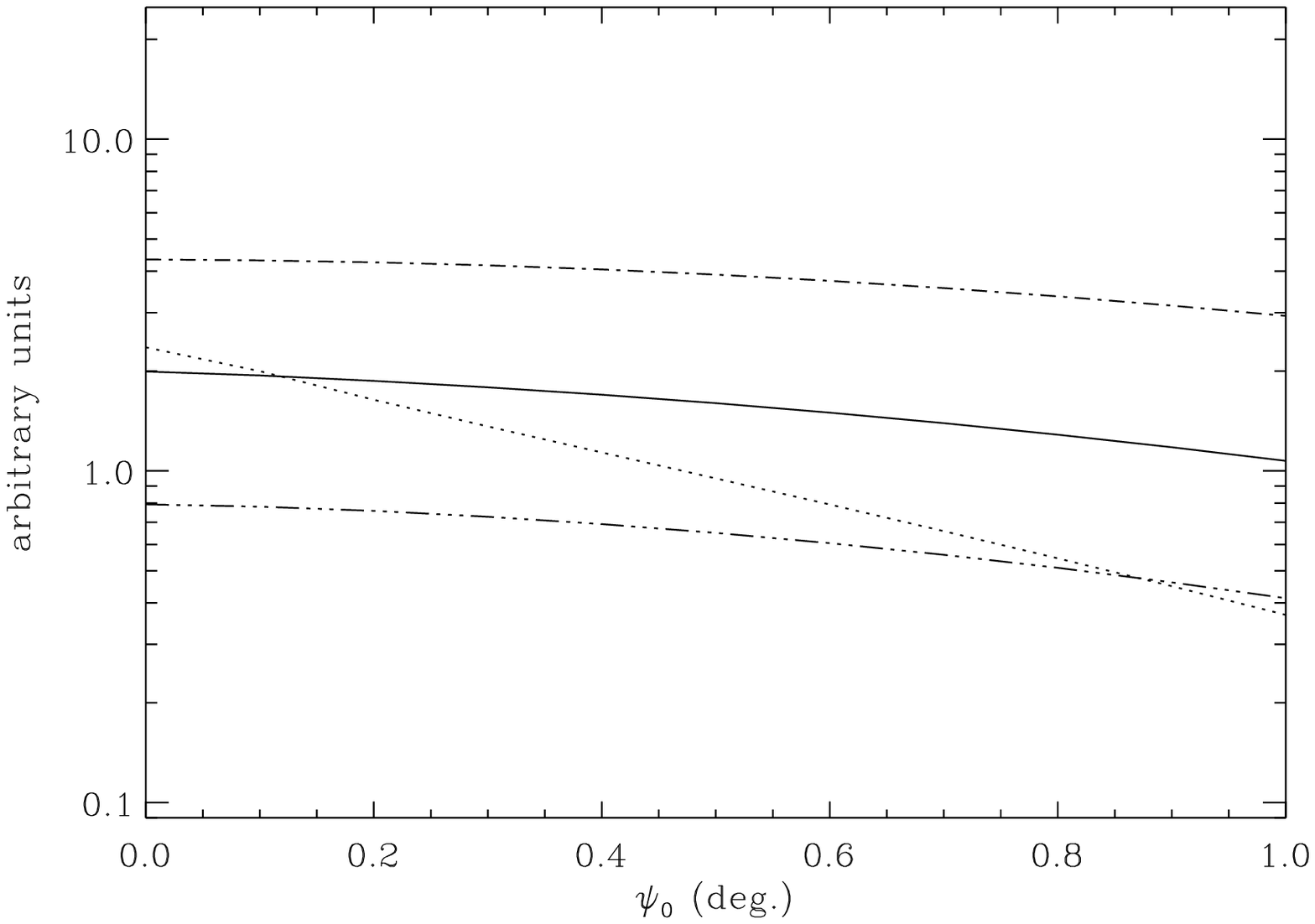}
  \end{minipage}% 
  \begin{minipage}[t]{0.5\linewidth}
    \centering
    \includegraphics[height=.25\textheight]{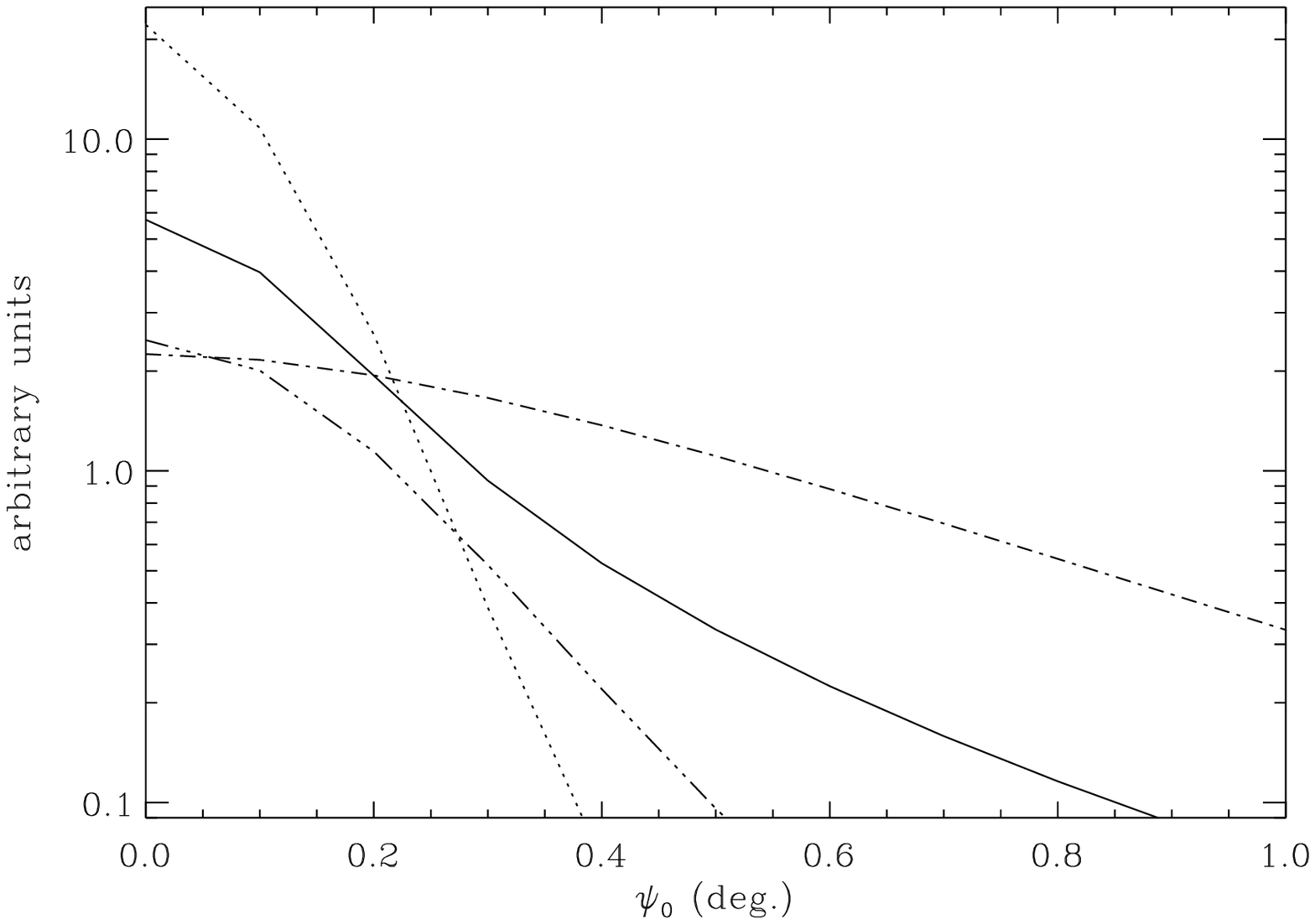}
  \end{minipage}
  \caption{Flux profiles for the four models of dark matter density profile used in \citep{profumo&kamion}, computed using the CACTUS PSF$=1\grado$ (left panel), and using an improved PSF$=0.1\grado$ (right panel).}  
    \label{cactuspsfs}
\end{figure}

As we can see in the left panel of Figure \ref{cactuspsfs}, if we adequately introduce this effect, it will be impossible to discriminate between different flux profiles, i.e. between the four models for the density profile described in \citep{profumo&kamion}, using the CACTUS PSF. Only the absolute flux could give us a clue, but as it was mentioned before, there are too many uncertainties in the \rm{y} axis. In the right panel of Figure \ref{cactuspsfs} the same exercise was done, but with an improved PSF$=0.1\grado$ (e.g. the MAGIC PSF). In this case we could distinguish between different flux profiles, i.e. different models for the dark matter density profile.

\section{Draco and GAW}

GAW is a R\&D path-finder experiment, under development, for wide field $\gamma$-ray astronomy. GAW will operate above 0.7 TeV and will have a PSF $\sim 0.2\grado$. It will consist of three identical telescopes working in stereoscopic mode (80m side). The mean goal of GAW is to test the feasibility of a new generation of IACTs, which join high sensitivity with large field of view. GAW is planned to be located at Calar Alto Observatory (Spain) and a first part of the array should be completed and operative within winter 2008. GAW is a collaborative effort of Research Institutes in Italy, Portugal, and Spain.

It is possible to make some calculations concerning the possibility to observe a $\gamma$-ray excess in the direction of Draco by GAW. If we are only interested in flux detectability (i.e. no discrimination between different dark matter density profiles), we should calculate simply the integrated flux that we will observe in GAW due to dark matter annihilation in Draco using Eq.(\ref{eq1}) with $E_{th}=0.7~TeV$. To make the calculations, we suppose Draco to be around $1.5\grado$ in the sky, a GAW PSF$=0.2\grado$ and a $S/N>5$. There are, however, large uncertainties (e.g. the exact value of $f_{SUSY}$ for 700 GeV, the Draco inner profile, etc). If we take $f_{SUSY} = 10^{-34}~ph \cdot GeV^{-2} \cdot cm^{-3} \cdot s^{-1}$ for 0.7 TeV (as in \citep{Prada04}), and a density profile given by Eq.(\ref{kazantzidis}) with $\alpha=1$, we obtain for the total flux:

\begin{equation}
F_{Draco} = 2.4876 \cdot 10^{-11}~ph~cm^{-2}~s^{-1}
\label{eqGAW2}
\end{equation}

\noindent Given this value, it may be possible to detect a $\gamma$-ray excess from Draco by GAW, since the minimum detectable flux in 50 hours at $5\sigma$ level with this IACT is $3.5\cdot 10^{-12}~ph~cm^{-2}~s^{-1}$ (\citep{bluebook}).

\section{Conclusions}

We can summarize the main conclusions as follows:
\begin{itemize}
\item The PSF of the instrument is crucial to estimate correctly the expected gamma flux profile due to dark matter annihilation in IACTs. 
\item The effect of the PSF could make it impossible to discriminate between different models of the dark matter density profile.
\item Concerning the $\gamma$-ray excess reported by CACTUS in the direction of Draco, their results (if real) should be interpreted carefully. There is no possibility to say, given the poor PSF of the experiment, if the dark matter density profile is cusp or core. May be they only detected an excess signal, which has to be confirmed.
\item Moreover, if the CACTUS excess is real, MAGIC should see the signal without problems and could distinguish among the different dark matter density profiles.
\end{itemize}

There is no doubt that GLAST, with a PSF$<0.1\grado$, will be very important in the indirect search of dark matter. In that sense, also IACTs with large field of view and high sensitivity are the next step in this search. The R\&D experiment GAW represents a first attempt in that direction.

%%%%%%%%%%%%%%%%%%%%%%%%%%%%%%%%%%%%%%%%%%%%%%%%
%% BACKMATTER
%%%%%%%%%%%%%%%%%%%%%%%%%%%%%%%%%%%%%%%%%%%%%%%%

\begin{theacknowledgments}
M.A.S.C. wants to specially thank M.C. S\'anchez-Gil for her help in mathematical issues. M.A.S.C. acknowledges the support of an I3P-CSIC fellowship in Granada. M.A.S.C. and F.P. also acknowledge the support of the Spanish AYA2005-07789 grant. This work was partially supported by the Polish Ministry of Science and Higher Education under grant 1P03D02726 and the Polish-Spanish exchange program of CSIC/PAN.
\end{theacknowledgments}

%bibliographystyle{aipproc}   % if natbib is available
\bibliographystyle{aipprocl} % if natbib is missing

\end{document}